\documentclass[twocolumn,twoside,slac_two]{revtex4}
\usepackage{graphicx}
\usepackage{fancyhdr}
\begin{document}

\title{Measurement of the Z boson transverse momentum spectrum on ATLAS with early data}
\author{L. Kashif}
\affiliation{Department of Physics, Harvard University, Cambridge, MA 02138, USA}
\begin{abstract}
One of the benchmark analyses to be performed with the first data at the CERN Large Hadron Collider will be the measurement of the \textit{Z} boson transverse momentum spectrum. In this talk, I present a prospective analysis for this measurement in the dimuon channel on the ATLAS experiment. The analysis uses simulated datasets at a center-of-mass energy of 10 TeV. After summarizing the motivations for the measurement, I discuss the \textit{Z} boson selection criteria, possible physics backgrounds, and background removal techniques with a focus on data-driven background determination. I then present my results, and conclude with an outlook toward the collision data expected in 2009-10.\end{abstract}

\maketitle

\thispagestyle{fancy}

\section{Introduction}
The first run of the Large Hadron Collider (LHC) at CERN is scheduled to start very soon. A number of benchmark analyses are being prepared by ATLAS (A Toroidal LHC Apparatus) and CMS (Compact Muon Solenoid), the two general-purpose experiments at the LHC. These analyses will help commission the detectors as well as establish known Standard Model physics in the new energy regime. 

One of the benchmark analyses is the measurement of the transverse momentum spectrum of the \textit{Z} boson. The Harvard group is developing an analysis to perform this measurement, which is presented in this article. In Section~\ref{sec:motiv}, we summarize the motivations for the measurement. In Section~\ref{sec:ana}, we give the salient details of our analysis, including data-driven estimations of the major backgrounds and background removal. Section~\ref{sec:result} presents the results of the analysis. We conclude in Section~\ref{sec:conc}.

\section{Motivations for measuring the \textit{Z} $p_T$ spectrum}
\label{sec:motiv}

The transverse momentum spectrum of the \textit{Z} boson is important for a number of reasons, both theoretical and practical. The principal among these are briefly discussed below.

\begin{figure}[h]
 \begin{center}
  \includegraphics[angle=270,width=80mm]{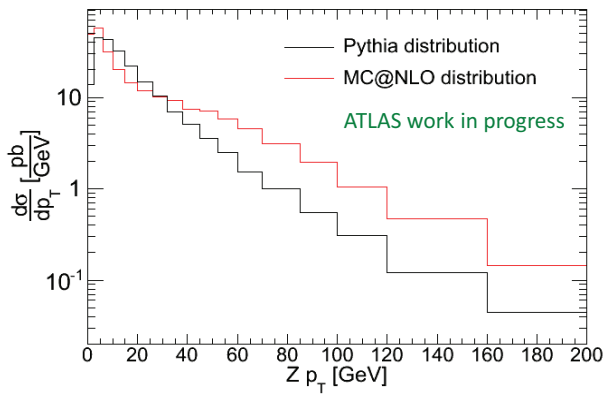}
 \end{center} 	
  \caption{\textit{Z} differential cross-section vs $p_{T}$ from two different MC generators: Pythia \textit{(black)}, which uses leading-order matrix elements, and MC@NLO \textit{(red)}, which uses next-to-leading order matrix elements.} 
   \label{fig:xsec_pt_pythia_mcatnlo}
\end{figure}

\begin{itemize}
	\item Test of QCD predictions: The high-$p_{T}$ region of the \textit{Z} $p_{T}$ spectrum is affected by perturbative QCD corrections, while non-perturbative corrections modify the low-$p_{T}$ region. As an example, Figure~\ref{fig:xsec_pt_pythia_mcatnlo} shows the \textit{Z} differential cross-section as a function of the \textit{Z} $p_{T}$ from two Monte Carlo programs: Pythia and MC@NLO. Pythia uses leading-order matrix elements, while MC@NLO uses next-to-leading-order matrix elements. The difference in the two distributions is clearly visible. 
	
Recently, a number of Monte Carlo tools have become available which combine electroweak and QCD corrections. Predictions from the state-of-the-art in theory can be tested by comparison with the measured \textit{Z} $p_{T}$ distribution.
	
	\item Discovery physics: Various physics scenarios beyond the Standard Model predict heavy particles that couple to the electroweak sector, and therefore can decay to high-$p_{T}$ \textit{Z} bosons. These include technicolor condensates, squarks and gluinos from supersymmetric models, right-handed quarks etc. If any of these scenarios is realized in nature, the decay signature of the heavy fields may well show up in the \textit{Z} $p_{T}$ spectrum.
	
	\item Inferring properties of the $Z \rightarrow \nu\nu$ decay: The invisible $Z \rightarrow \nu\nu$ decay is a background to many processes involving missing transverse energy, such as Supersymmetry and extra-dimension models where a graviton escapes into the higher-dimensional 'bulk'. The $Z \rightarrow \mu\mu$ decay can be used to indirectly measure the rate and properties of the invisible mode. In order to do this, we need to measure the $p_{T}$ spectrum of \textit{Z} bosons in the muon channel.
	
	\item Both the ATLAS and CMS experiments will eventually use the total \textit{Z} production cross-section to monitor luminosity on a run-by-run basis. Before we can do that, however, we must understand the \textit{Z} $p_{T}$ spectrum very well, especially the low-$p_{T}$ region.
	
\end{itemize}	
	
\section{Details of the analysis}
\label{sec:ana}

Our analysis is based on official ATLAS Monte Carlo datasets corresponding to a center-of-mass energy of 10 TeV. The datasets were generated with the Pythia and MC@NLO generators\footnote{The PHOTOS tool was interfaced with both Pythia and MC@NLO to generate final-state photon radiation.}, and were fully simulated using the GEANT4 ATLAS detector simulation. The sample we use for the signal channel corresponds to an integrated luminosity of 40 $pb^{-1}$, which is a realistic estimate of the amount of data expected in the first year of LHC running.

Table~\ref{tab:z_cuts} summarizes our \textit{Z} boson selection criteria. 

\begin{table}[htbp]
  \caption{\label{tab:z_cuts} The criteria used to select events likely to contain \textit{Z} bosons. The fraction of events passing each cut is shown on the right. Errors are statistical only. } 
  \begin{center}
    \begin{tabular}{|l|c|} \hline
    Selection criterion & Events passing cut (\%) \\ \hline
		20 GeV single muon trigger & 69.7 $\pm$ 0.4 \\ \hline
		At least one $\mu^{+}$ and one $\mu^{-}$ & 49.8 $\pm$ 0.3 \\
		reconstructed & \\ \hline
		$|\eta| < 2.5$ for both muons & 47.1 $\pm$ 0.3\\ \hline
		$p_T > 20$ GeV for both muons & 41.6 $\pm$ 0.3\\ \hline
		76 GeV $<M_{\mu\mu}<$ 106 GeV & 38.5 $\pm$ 0.3\\ \hline
		\end{tabular}	
  \end{center}
\end{table}

\subsection{Background reduction}

Possible physics backgrounds to the signal channel include:

\begin{itemize}
 \item $W \rightarrow \mu\nu + jet$
 \item $b \bar{b} \rightarrow \mu\mu + X$
 \item $t \bar{t}$ dimuon decays
 \item $Z \rightarrow \tau\tau \rightarrow \mu\nu_{\mu}\nu_{\tau}$ $\mu\nu_{\mu}\nu_{\tau}$
 \item $WW, ZZ, WZ$
\end{itemize} 

The first three channels have the largest cross-sections, which are the ones we are studying at this time. We estimate our backgrounds using Monte Carlo samples. However, we are investigating methods to extract the \textit{W} and $b \bar{b}$ backgrounds from data, as we will discuss shortly. To minimize backgrounds, we use two isolation criteria defined in a cone of size $\Delta R = 0.4$ around the muon track, where

\begin{equation}
	\Delta R = \sqrt{\Delta\eta^2 + \Delta\phi^2}
\end{equation}

and $\eta$ and $\phi$ are the pseudorapidity and the azimuthal coordinate respectively. The isolation criteria are:

\begin{itemize}
 \item number of tracks in cone, and
 \item the total transverse momentum of tracks, $\Sigma p_T$, in cone.
\end{itemize} 

\begin{figure*}[ht]
 \begin{center}
  \includegraphics[angle=270,width=150mm]{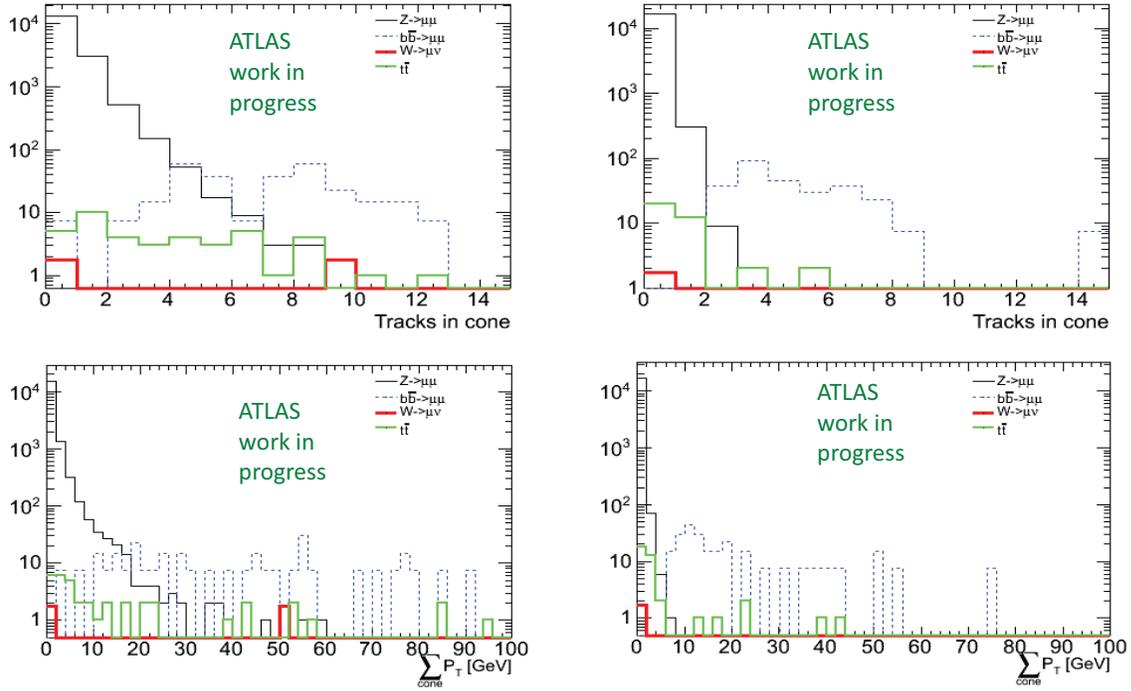}
 \end{center} 	
  \caption{Number of tracks \textit{(top)} and $\Sigma p_T$ of tracks \textit{(bottom)} in cone of size \textit{R} = 0.4. Distributions for the muon with the larger value of the quantity is shown on the left, and those for the muon with the smaller value on the right.} 
   \label{fig:Ntrack_trackPt_sig_bg}
\end{figure*}

\begin{figure*}[ht]
 \begin{center}
  \includegraphics[angle=270,width=150mm]{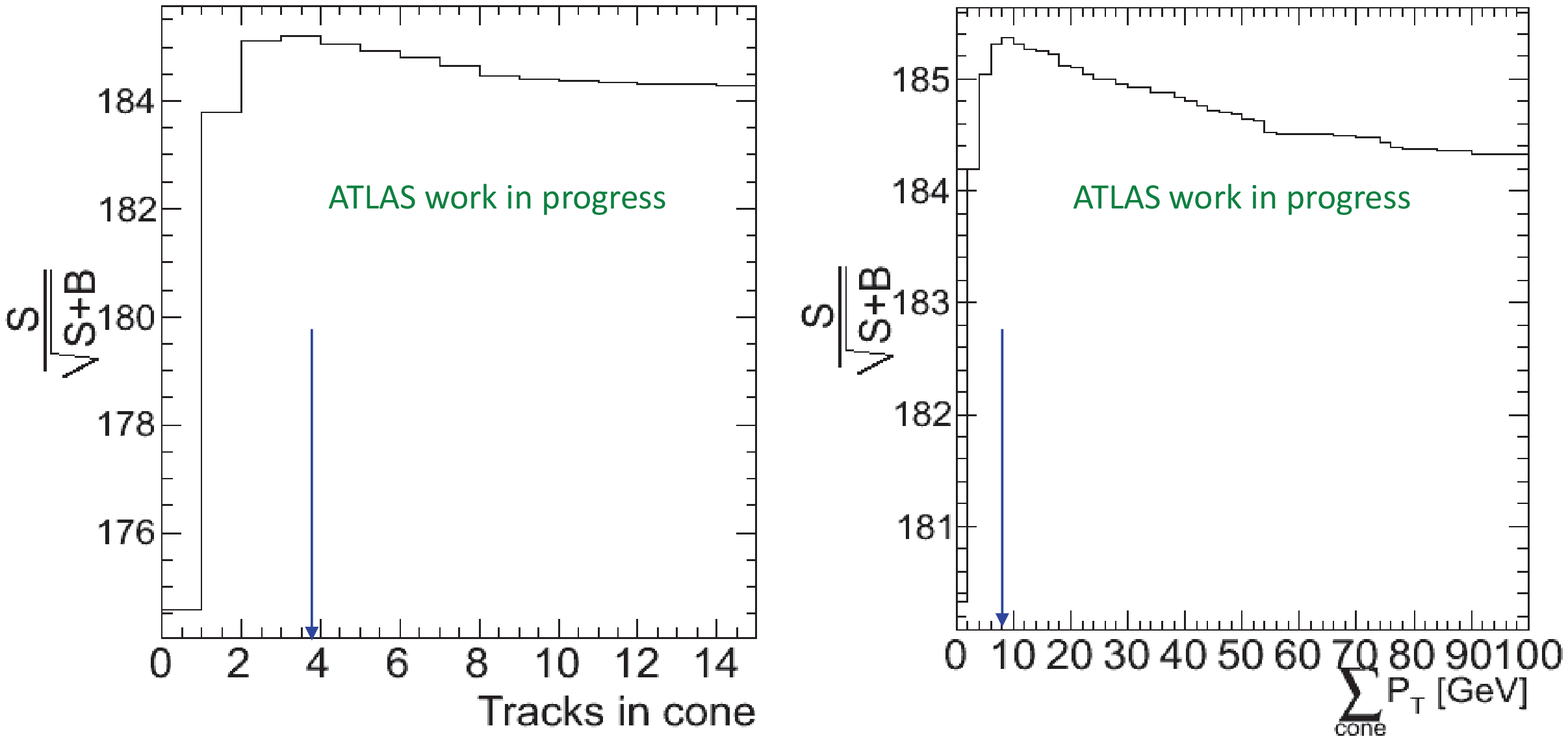}
 \end{center} 	
  \caption{Optimization of the muon isolation cuts: \textit{(left)} number of tracks in cone of size \textit{R} = 0.4 around the muon track, and \textit{(right)} $\Sigma p_T$ of tracks in cone. On each plot, the arrow indicates the value of the quantity where the peak in the distribution occurs.} 
   \label{fig:Ntrack_trackPt_sig_bg_opt}
\end{figure*}

Figure~\ref{fig:Ntrack_trackPt_sig_bg} shows the distribution of these variables for the signal channel and the three main background channels. We optimize the cuts on these quantities using distributions of $\frac{S}{\sqrt{S+B}}$, where\textit{S} and \textit{B} refer respectively to the signal and the total contribution from all three background channels. Figure~\ref{fig:Ntrack_trackPt_sig_bg_opt} shows the $\frac{S}{\sqrt{S+B}}$ distributions for the two isolation criteria.

The optimized values of the cuts are given by the position of the peaks:

\begin{itemize}
 \item number of tracks in cone $<$ 4,
 \item $\Sigma p_T$ of tracks in cone $<$ 8 GeV.
\end{itemize} 

\begin{figure*}[t]
 \begin{center}
  \includegraphics[angle=270,width=150mm]{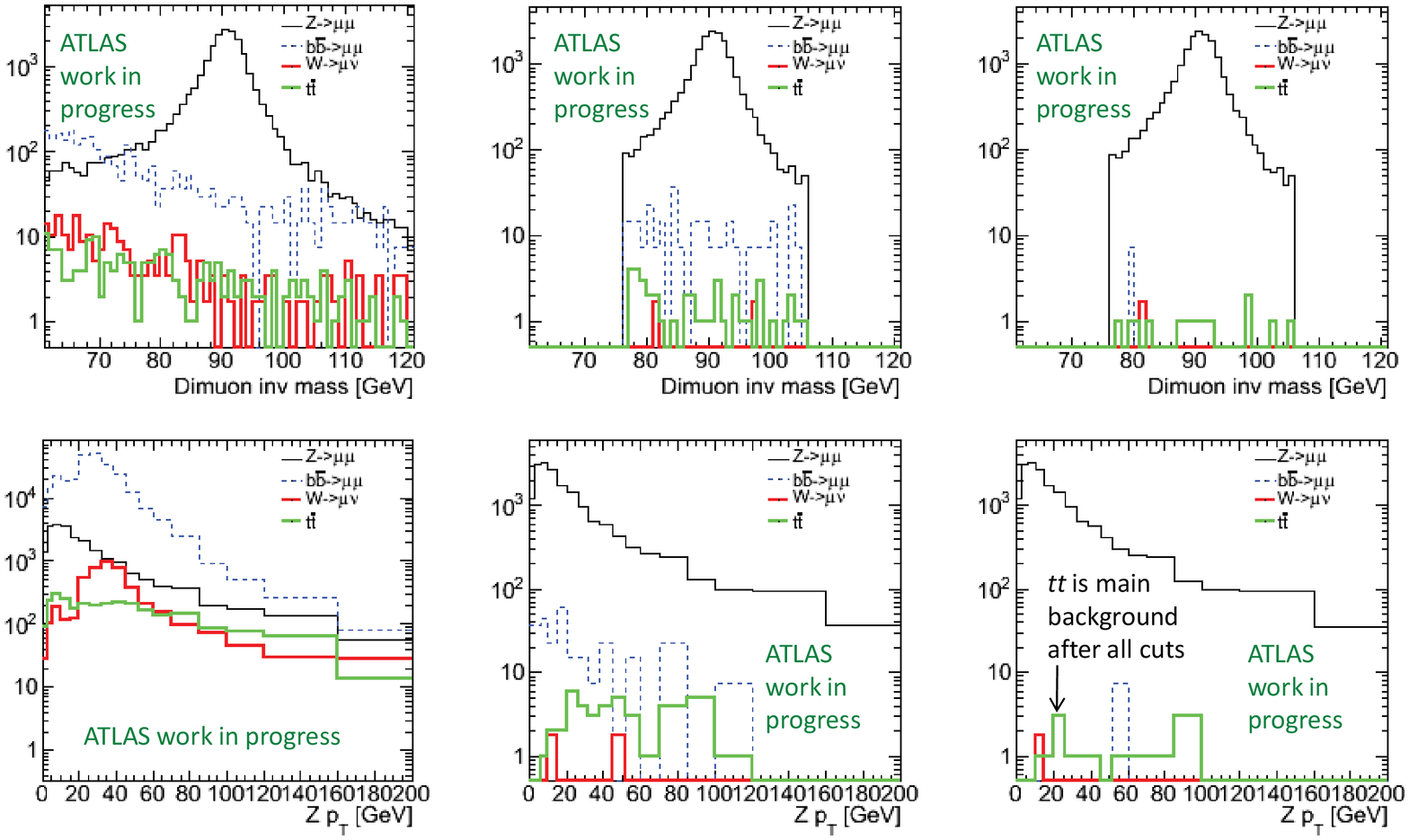}
 \end{center} 	
  \caption{Distribution of signal and background events in dimuon invariant mass \textit{(top)} and dimuon $p_T$ \textit{(bottom)} distributions. \textit{Left}: before cuts, \textit{middle}: after Z selection cuts, \textit{right}: after isolation cuts.} 
   \label{fig:invmass_pt_sig_bg_cuts}
\end{figure*}

Figure~\ref{fig:invmass_pt_sig_bg_cuts} shows the distribution of the dimuon invariant mass and the dimuon $p_T$ spectra from the signal channel and the three background channels. Distributions for the quantities are shown before any cuts, after the \textit{Z} boson selection cuts and before the isolation cuts, and after the isolation cuts. $t\bar{t}$ dimuon decays are the largest background contribution after all cuts. This is expected, since muons from this source are very similar to those from \textit{Z} decays. Following the isolation cuts, the remaining total background contamination is $\approx 0.2\%$ of the signal, which is small enough that we ignore it in the remainder of the analysis.

\subsection{Estimation of $W \rightarrow \mu\nu$ background from data}

In a fraction of the $W \rightarrow \mu\nu$ events, the muon from the \textit{W} decay combines with a muon in a jet to mimic the $Z \rightarrow \mu\mu$ signal. To estimate this background in data, we make the following assumption:

The fraction of $Z \rightarrow \mu\mu$ events in which three muons pass our selection and isolation cuts is equal to the fraction of $W \rightarrow \mu\nu$ events in which two muons pass the same cuts.

To implement this assumption in a data sample, we will first apply the selection and isolation cuts on the sample, and then count the number of events in which three muons pass all the cuts. Multiplying this number by the ratio of the total \textit{W} cross-section to the total \textit{Z} cross-section will give an estimate of the $W \rightarrow \mu\nu$ events remaining after the cuts.

We tested the validity of our assumption on the Monte Carlo samples. In the signal sample, 0.029 $\pm$ 0.008\% events (13 events) have a third muon passing all the cuts, while 0.027 $\pm$ 0.003\% events in the $W \rightarrow \mu\nu$ sample have a second muon passing the cuts. These numbers are consistent with each other, suggesting that our assumption is valid.

\subsection{Estimation of $b\bar{b}$ background from data}

\begin{figure}[h]
 \begin{center}
  \includegraphics[angle=270,width=80mm]{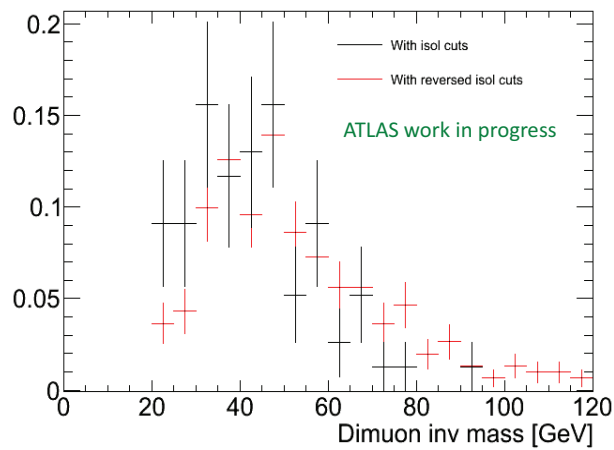}
 \end{center} 	
  \caption{Dimuon invariant mass distribution in $b\bar{b}$ events with muon isolation cuts \textit{(black)} and with reversed isolation cuts \textit{(red)}, normalized to unit area.} 
   \label{fig:bb_invmass_isol_nonisol_norm}
\end{figure}

In this case, the idea is to use a sample of $b\bar{b}$ events with non-isolated muons to estimate the number of $b\bar{b}$ events remaining after isolation cuts. In other words, we want to use a sample of $b\bar{b}$ events with non-isolated muons as a template for $b\bar{b}$ events with isolated muons. Figure~\ref{fig:bb_invmass_isol_nonisol_norm} shows the dimuon invariant mass distribution in $b\bar{b}$ events with our isolation cuts and with reversed isolation cuts. Reversing the isolation cuts on a data sample selects a very pure sample of $b\bar{b}$ events with non-isolated muons. Within statistics, the two invariant mass distributions have similar shape, indicating that the distribution with non-isolated muons can be used as a template for $b\bar{b}$ background after isolation cuts. The signal template is a dimuon invariant mass distribution from $Z \rightarrow \mu\mu$ events after isolation cuts.A template fit now gives the fractions in which the signal and background templates must be combined to obtain the observed distribution.

\begin{figure*}[ht]
 \begin{center}
  \includegraphics[angle=270,width=150mm]{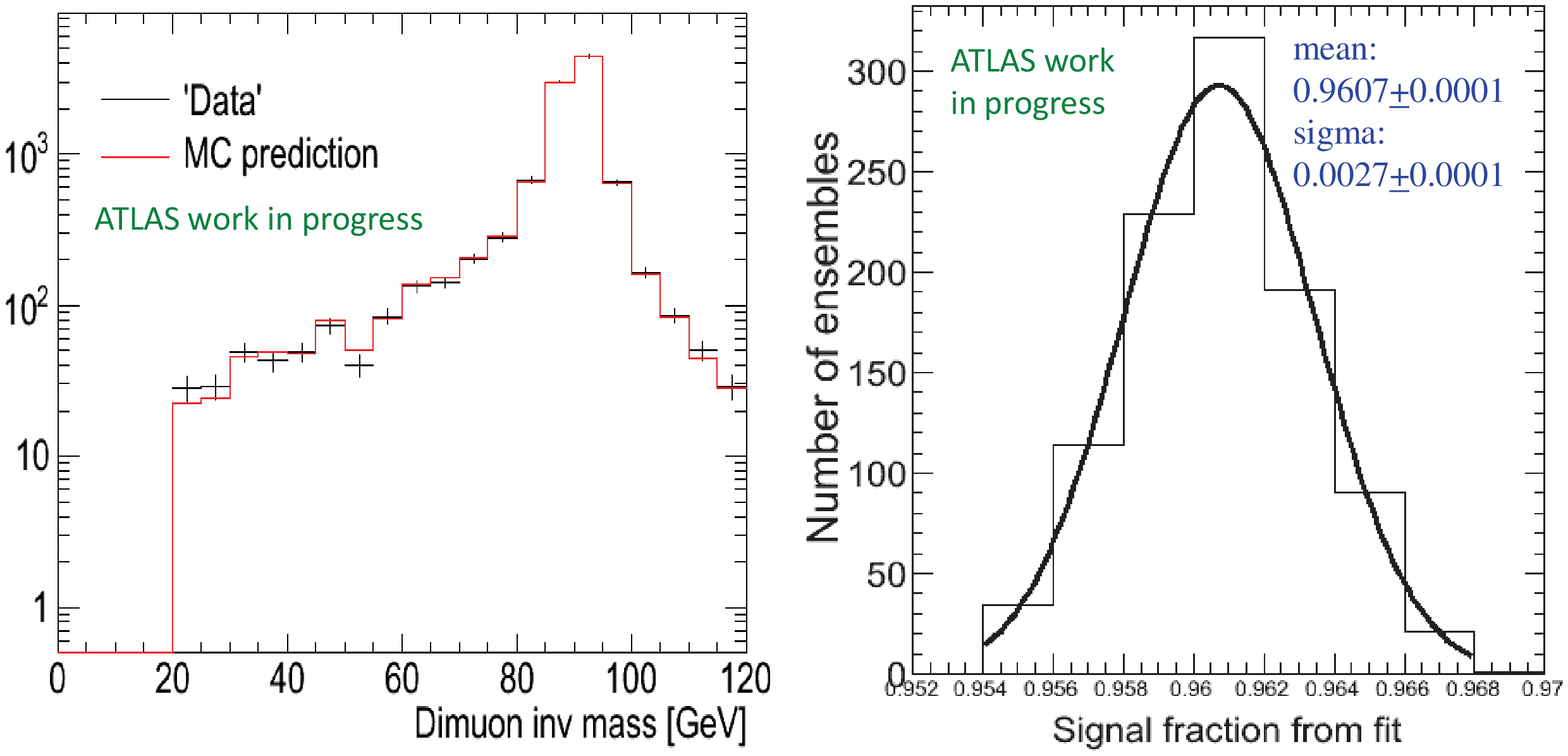}
 \end{center} 	
  \caption{\textit{(Left)} Dimuon invariant mass distribution from 'data'  and from the template fit result. \textit{(Right)} \textit{Z} fractions from the template fit in 1000 pseudo-experiments, in which the content of each 'data' bin was varied using Poisson fluctuation.} 
   \label{fig:template_fit}
\end{figure*}

We test this method on our Monte Carlo samples. The template fit gives 96 $\pm$ 1\% for signal fraction and 3.9 $\pm$ 0.4\% for $b\bar{b}$ background fraction. The known inputs to our 'data' distribution were 97\% signal and 3\% $b\bar{b}$ events, so the template fit seems to yield the correct numbers. The left-hand plot in Figure~\ref{fig:template_fit} shows the dimuon invariant mass distributions from 'data' and from the template fit result for invariant mass $>$ 20 GeV; good agreement is observed. The right-hand plot in Figure~\ref{fig:template_fit} shows the distribution of signal fraction from the template fit in 1000 pseudo-experiments. In each experiment, the numbers of signal and $b\bar{b}$ events in each bin of the 'data' distribution were varied within their Poisson fluctuations, and the template fit was done using the same signal and $b\bar{b}$ templates. The width of the Gaussian fit gives an estimate of the resolution that can be expected from the template fit.

\section{Results}
\label{sec:result}

The measured \textit{Z} $p_T$ spectrum must be corrected for various detector and reconstruction effects, including:

\begin{itemize}
 \item trigger and reconstruction efficiencies,
 \item geometric and kinematic acceptance,
 \item resolution smearing.
\end{itemize}
 
\begin{figure*}[ht]
 \begin{center}
  \includegraphics[angle=270,width=150mm]{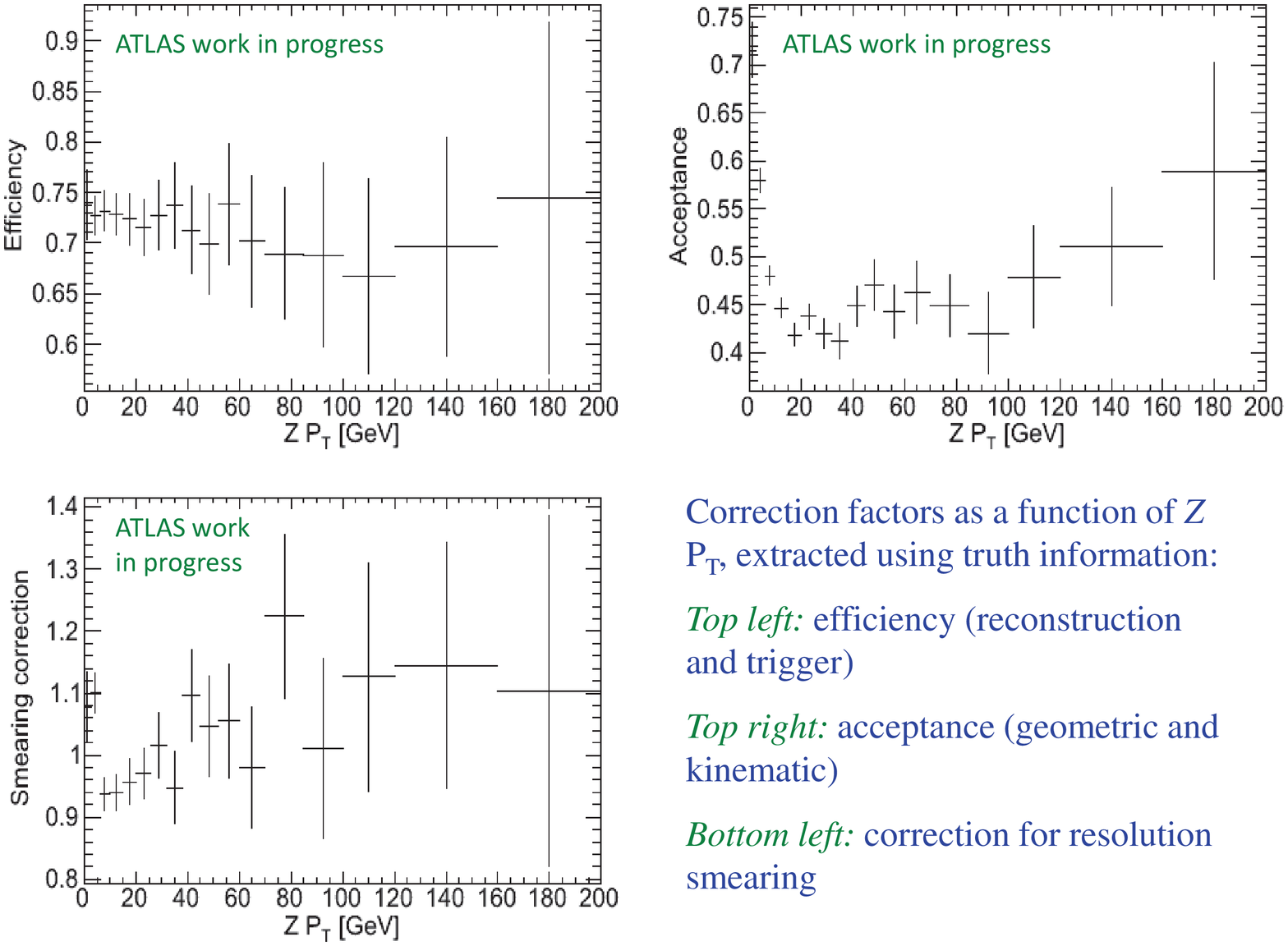}
 \end{center} 	
  \caption{Corrections to the \textit{Z} $p_T$ spectrum.} 
   \label{fig:corrections}
\end{figure*}

\begin{figure}[h]
 \begin{center}
  \includegraphics[angle=270,width=80mm]{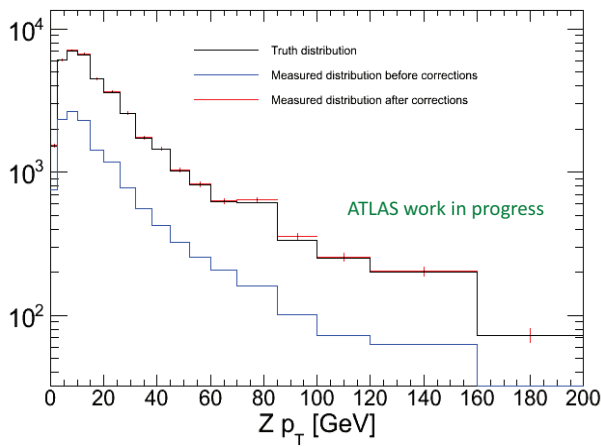}
 \end{center} 	
  \caption{\textit{Z} $p_T$ distribution from Monte Carlo truth \textit{(black)}, measured (uncorrected) \textit{(blue)} and corrected \textit{(red)}. The plot corresponds to 40 $pb^{-1}$ of integrated luminosity. Errors are statistical.} 
   \label{fig:z_pt_dist}
\end{figure}

We derive the correction factors in \textit{Z} $p_T$ bins using Monte Carlo truth information (Figure~\ref{fig:corrections})\footnote{With collision data, reconstruction and trigger efficiencies will be estimated from data, most likely using a tag-and-probe approach.}. Figure~\ref{fig:z_pt_dist} shows the measured \textit{Z} $p_T$ spectrum from our sample before and after the above corrections, as well as the true spectrum. Since the correction factors in this case were obtained using the true spectrum, the good agreement between the corrected measurement and the prediction is, of course, expected.

\begin{figure}[h]
 \begin{center}
  \includegraphics[angle=270,width=80mm]{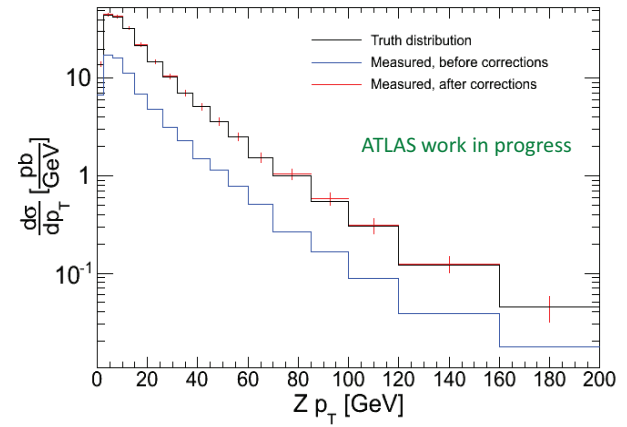}
 \end{center} 	
  \caption{\textit{Z} differential cross-section vs $p_T$ from Monte Carlo truth \textit{(black)}, measured (uncorrected) \textit{(blue)} and corrected \textit{(red)}. The plot corresponds to 40 $pb^{-1}$ of integrated luminosity. Errors are statistical.} 
   \label{fig:z_xsec_vs_pt}
\end{figure}

The differential cross-section as a function of \textit{Z} $p_T$ is given by:

\begin{equation}
	\frac{d\sigma}{dp_{T}} = c_{i}\frac{s_{i}-b_{i}}{\Delta p_{T_{i}}L_{int}\epsilon_{i}A_{i}}
\end{equation}

where 

$s_i$  = signal events in \textit{i}-th $p_T$ bin,

$b_i$  = background events in \textit{i}-th $p_T$ bin,

$\Delta p_{T_{i}}$ = width of \textit{i}-th $p_T$ bin,

$L_{int}$ = integrated luminosity,
 
$A_i$   = acceptance for \textit{i}-th bin,

$\epsilon_i$ = efficiency for \textit{i}-th bin,

$c_i$ = smearing correction for \textit{i}-th bin. 

Figure~\ref{fig:z_xsec_vs_pt} shows the \textit{Z} differential cross-section before and after corrections, together with the expected distribution from Monte Carlo truth. As before, the good agreement between the corrected measurement and the prediction is expected.

\section{Conclusion and outlook}
\label{sec:conc}

We are developing an analysis to measure the \textit{Z} $p_T$ spectrum in the dimuon channel on the ATLAS experiment. During the first LHC run, we expect $\approx 100 pb^{-1}$ of delivered integrated luminosity, so that we should have a few tens of $pb^{-1}$ of luminosity on tape, assuming realistic data-taking efficiency. Our analysis is geared toward that amount of data. We are exploring methods to extract the major backgrounds to our signal channel with minimal dependence on Monte Carlo. 

In the first run, the instantaneous luminosity will be in the range $10^{29-31}$ $cm^{-2}s^{-1}$, so that cavern background and event pileup are not likely to be major issues. The main challenge for this analysis will be understanding the detector, including muon trigger acceptance and efficiency, offline reconstruction efficiency, muon momentum scale and resolution, and the systematics associated with all of them.



\end{document}